\documentstyle[tighten,aps,epsf]{revtex} 

\begin{document}

\draft

\title{Chaos in coplanar classical collisions with particles interacting 
through $ r^{-2}$ forces}

\author{Fabio Sattin$^{(1)}$\thanks{E--mail: sattin@igi.pd.cnr.it} and 
Luca Salasnich$^{(2)}$\thanks{E--mail: salasnich@hpmire.mi.infm.it}}

\address{$^{(1)}$ Consorzio RFX and INFM, Unit\`a di Padova, \\ Corso 
Stati Uniti 4, I 35127 Padova, Italy \\ $^{(2)}$Istituto Nazionale per la 
Fisica della Materia, Unit\`a di Milano, \\ Dipartimento di Fisica, 
Universit\`a di Milano, \\ Via Celoria 16, I 20133 Milano, Italy \\ 
Istituto Nazionale di Fisica Nucleare, Sezione di Padova,\\ Via Marzolo 8, 
I 35131 Padova, Italy}

\maketitle

\abstract{The scattering among three particles interacting through $1/r^2$ 
forces, with opposite charges and widely different masses, is studied in a 
coplanar geometry. The present work shows that at low impact velocities 
the output of the collision presents typical fingerprints of chaos. The 
details of the process are investigated. } 


%\section{Introduction}

\par
Starting from the works of Gaspard and Rice \cite{gaspard}, chaotic 
scattering was found to be ubiquitous. A large amount of work has been 
done on potential scattering ({\it i.e.} scattering of an elementary 
particle on fixed potential centers) \cite{bleher,bleher2,troll,brandis}, 
but also the scattering among at least three interacting bodies has 
received great attention (see the work of Petit and H\'enon on the 
scattering between a planet and two satellites \cite{petit}, or the system 
helium nucleus plus two electrons \cite{yuan}). A monographic issue about 
this subject has been published in the journal CHAOS \cite{chaos}. 
\par
In all these studies the projectile and a target preserve their internal 
structure during the collision, but one can analyze also reactive 
collisions.
Kov\'acs and Wiesenfeld \cite{kovacs} studied, in a collinear geometry, 
the scattering between an atom and a diatomic molecule: $ A + BC 
\leftrightarrow ABC \leftrightarrow AB + C$. More recently, we studied the 
chaotic behavior in the reaction between a hydrogen atom (proton plus 
electron) and a projectile proton interacting through Coulomb forces 
\cite{sattin}. We analyzed the full three-dimensional problem at very low 
impact energies. We found that the transition from
regular to chaotic scattering appears when impact velocity $v_p$ is 
reduced to a value below about
$1/10$ of the classical electron velocity $v_e$. 
\par
In this paper we present another investigation on the same subject: we 
think that it is worth consideration since collisions with rearrangement 
between electrically charged particles are one current topic in atomic 
physics, both on the theoretical as well the experimental side
\cite{bransden}.  The phase space of a system of three particles moving in 
3D is too large to be easily handled; limiting to 2D will allow us to do a 
more detailed and accurate study.  We will address the following topics: a) 
Do features of chaotic scattering appear in two--dimensional collisions?  
b) If so, do they appear in the form of a sharp order--chaos transition, or 
as a smooth transition?  c) Is it possible to detect some traces of 
irregular behaviour also in the heavy particles motion, instead that only 
when looking at the lightest one?  d) Finally, an investigation of the 
dynamics of the electron during the scattering is done.  It gives an 
insight on how the discontinuities on the output function appear.

%\section{Theory}

The final state of the projectile may be defined through a set of 
parameters $\{ A_f \}$ ({\it e.g.} angle of deflection, final velocity, 
\ldots), which are functions of the input quantities $\{ A_i \}$ ({\it 
e.g.} impact parameter, impact velocity, \ldots). When the set of values 
$\{ A_f \}$ depends sensitively by $\{ A_i \}$, {\it i.e.} a finite 
variation of $\{ A_f \}$ is caused by an infinitesimal variation of $\{ 
A_i \}$, then the system is chaotic. \par
Usually, one studies the so--called {\it scattering functions}, which are 
the output variables as a function of only one input variable, keeping the 
other input variables fixed. If a scattering function is fractal then the 
system is chaotic but the converse is not necessarily true. In fact, Chen 
{\it et al.} \cite{chen} showed that time-independent Hamiltonian systems 
with more than two degrees of freedom can have chaotic sets with nonzero 
fractal dimension while at the same time the scattering functions do not 
show any fractal property. The scattering function has fractal behaviour 
only if the Hausdorff dimension $D_{c}$ of the chaotic invariant set 
satisfies the inequality 
% eq 1
\begin{equation}
D_{c} > 2 N - (2 + q)
\end{equation}
where $N$ is the number of degrees of freedom and $q$ is the number of 
conserved quantities of the system (in our case $N = 6$ and $q = 4$). 
Therefore a positive answer to the above question (a) will give us also a 
lower bound estimate of the Hausdorff dimension of the chaotic set of the 
system. 
\par
Our scattering problem has the following alternative outcomes: 
% eq 2,3,4
\begin{eqnarray}
H + H^+ &\to& H^+ + H \\
&\to& e + H^+ + H^+ \\
&\to& H + H^+ \quad .
\end{eqnarray}
At the end of the collision the electron may be captured by the projectile 
nucleus (charge transfer), may be ionized (ionization), or may remain 
bound to the target (excitation of the target). The values of the masses 
and charges are $m_{H^+} = 1836$, $m_e \equiv 1836$, $Q_e=-Q_{H^+}=-1$ 
(atomic units will be used throughout this paper). 
\par
In our geometry the scattering plane is the plane $(x,y)$. The target 
proton is initially fixed at the origin, ${\bf r}_t=(0,0)$, and at rest, 
${\bf v}_t=(0,0)$. The position of the electron is ${\bf r}_e=(r_e 
\cos{\theta + \omega_e t}, r_e \sin{\theta + \omega_e t})$, where $r_e$ is 
the radius, $\theta$ is the initial phase, $\omega_e = v_e/r_e$ is the 
angular velocity. Both $v_e$ and $r_e$ are equal to $1$ in our units. The 
projectile proton is prepared with velocity $v_p$ in the positive 
direction of the $y$-axis and at the position ${\bf r}_p=(b,-a)$, where 
$b$ is the impact parameter and $a>0$ must be chosen so large that the 
target and the projectile may be considered initially as non--interacting. 
In these simulations $a \simeq 10$ was found to be a good choice. 
\par
The total energy of the system is given by 
% eq 5
\begin{equation}
E= {1\over 2}m_{H^+}v_t^2 + {1\over 2}m_{H^+}v_p^2 + {1\over 2}m_e v_e^2 + 
{1\over |{\bf r}_{t}-{\bf r}_{p}|} - {1\over |{\bf r}_{t}-{\bf r}_e|} -
{1\over |{\bf r}_{p}-{\bf r}_e|} \; .
\end{equation}
The classical equations of motion are numerically integrated by using a 
variable--order variable--step Adams algorithm (routine NAG D02CJF).
\par
Please, note that the choice of the outcome is done on the basis of the 
energy of the electron with respect to the two bodies. An algorithm to 
identify all the possible outcomes is given in ref. \cite{tokesi}. 
In particular, in our system three parameters remain free: $\theta$, 
$v_p$, and $b$. Numerical simulations have been carried out varying all of them. 
The output variables $\{ A_f \}$ defined above reduce to just one element--the 
final state of the electron. It is a discrete function of its energy. Discrete 
variables are usually not employed in these studies but are not a novelty:
already Bleher {\it et al} \cite{bleher1} investigated a chaotic system 
with a finite number of possible outcomes. 

%\section{Results}

\par
In Figures 1 and 2 some consecutive blow-ups of the scattering function 
are presented as a function of $b$, for two different choices of $v_p$ and 
$\theta$. There are regions where the outcome is a regular function of 
$b$, alternated with other regions where an irregular behaviour appears. 
One recognizes a kind of self similarity, the same pattern repeating at 
smaller and smaller scales. This is the distinctive aspect of fractal 
behaviour. The same results appear varying $\theta$, at $v_p$ and $b$ 
fixed. 
\par
A clear picture is obtained by studying the scattering function {\it 
versus} the proton velocity $v_p$.
In Figure 3 this is shown for one couple of values $(b,\theta )$. It is 
visible the
great degree of irregularity at small $v_p$'s. For higher $v_p$'s the 
zones where the electron is captured and those where it remains close to 
the target proton are more clear--cut. Ionization is by far the least 
probable process. One may judge that the transition order--chaos does 
appear quite smooth and the scattering at $v \simeq 0.2, 0.3 $ is not yet 
entirely regular.
\par
In order to quantify the amount of chaos one can compute the fractal 
dimension of the scattering function. Given a reference trajectory with 
initial impact parameter $b$, we slightly modify the impact parameter to 
$b +\epsilon$ and make another run. The results of the two runs are 
compared. The starting trajectory is said uncertain if the two final 
states are different. The fraction $f(\epsilon )$ of uncertain 
trajectories as a function of the parameter $\epsilon$ scales as 
% eq 6
\begin{equation}
f(\epsilon) \sim \epsilon^{1-d} \; , \;\;\;\;\;\; \hbox{for} \;\; \epsilon 
\to 0 \; ,
\end{equation}
where $d$ is the capacity dimension \cite{bleher2}. In Figure 4 we plot 
the capacity
dimension $d$ as a function of $v_p$. There is a regular decrease of $d$ 
for $v_p < 0.3$, then $d$ is nearly constant but positive. 
\par
One of the main purposes of this work is to analyze in detail the 
behaviour of the electron in correspondence of a singularity of the 
scattering function. In Figure 5 the trajectories of the particles are 
shown for two almost identical runs, differing only by a very small value 
of the impact parameter.
We see that initially the electron trajectories in the two cases nearly 
overlap. The two paths begin to differ when the nuclei reach the point of 
closest approach. The potential felt by the electron may be sketched as 
two wells located around the nuclei divided by a symmetrical ridge. 
Obviously, the location of the ridge dynamically evolves in time as the 
nuclei move. The electron initially lies in one of the two potential wells 
and it is able to escape to the other well only if at some time during the 
collision it passes near the ridge with enough velocity to cross it. Note 
that we are referring to the component of the velocity orthogonal to the 
potential ridge. Once the electron has fallen into the other potential 
well it is very unlikely that it can cross the ridge the opposite way and 
come back to the former nucleus. Two electron differing by an 
infinitesimal value of the velocity, $u, u + \epsilon$ will undergo 
entirely different fates provided that $u + \epsilon$ is not enough to 
cross the potential barrier while $u$ is. 
\par
It is interesting to notice that chaotic features appear only as 
singularities of the scattering function: An analysis of the delay time 
(the time needed by the electron to quit the interaction region) reveals 
that it is not a sensitive function of the scattering. This is related to 
the fact that the electron does closely follow one of the nuclei, and the 
trajectories of the heavy particles are always regular: no complicated 
patterns appear from a study of their behaviour.

%\section{Summary}

\par
In summary, a study of the two--dimensional low energy scattering between 
charged particles has been carried out. Simulations have been done using a 
larger number of runs than in our previous paper \cite{sattin} so 
diminishing statistical errors. The results confirm the findings of ref. 
\cite{sattin} that the scattering becomes irregular when the energy is low 
enough. An insight on why chaos appears has been given by looking at the 
detailed trajectories of the electron. Abrupt discontinuities in the 
output function are correlated with the structure of the potential seen by 
the electron. \par
It is useful to spend some words about the validity of the model used. At 
very small impact velocities, it is not entirely justified to treat the 
heavy particles as classical objects, and even less the light particle. 
The processes (2-4) should be studied within the framework of quantum 
mechanics. However: i) one may imagine to apply the same apparatus not to 
the ground state but to a Rydberg state, where using the classical 
mechanics is justified; ii) the use of classical or semiclassical models 
has recently been extended with satisfactory results in realms thought 
previously to be amenable only to quantum treatment, see for example the 
classical description of the helium atom \cite{yamamoto}, or the treatment 
of the hydrogen ionization by electron scattering at energies near 
threshold \cite{rost1}. In that work, furthermore, the evidence for chaos 
was found, reminiscent of the results shown in this paper.

% fig 1
\begin{figure}
\epsfxsize=12cm
\epsfbox{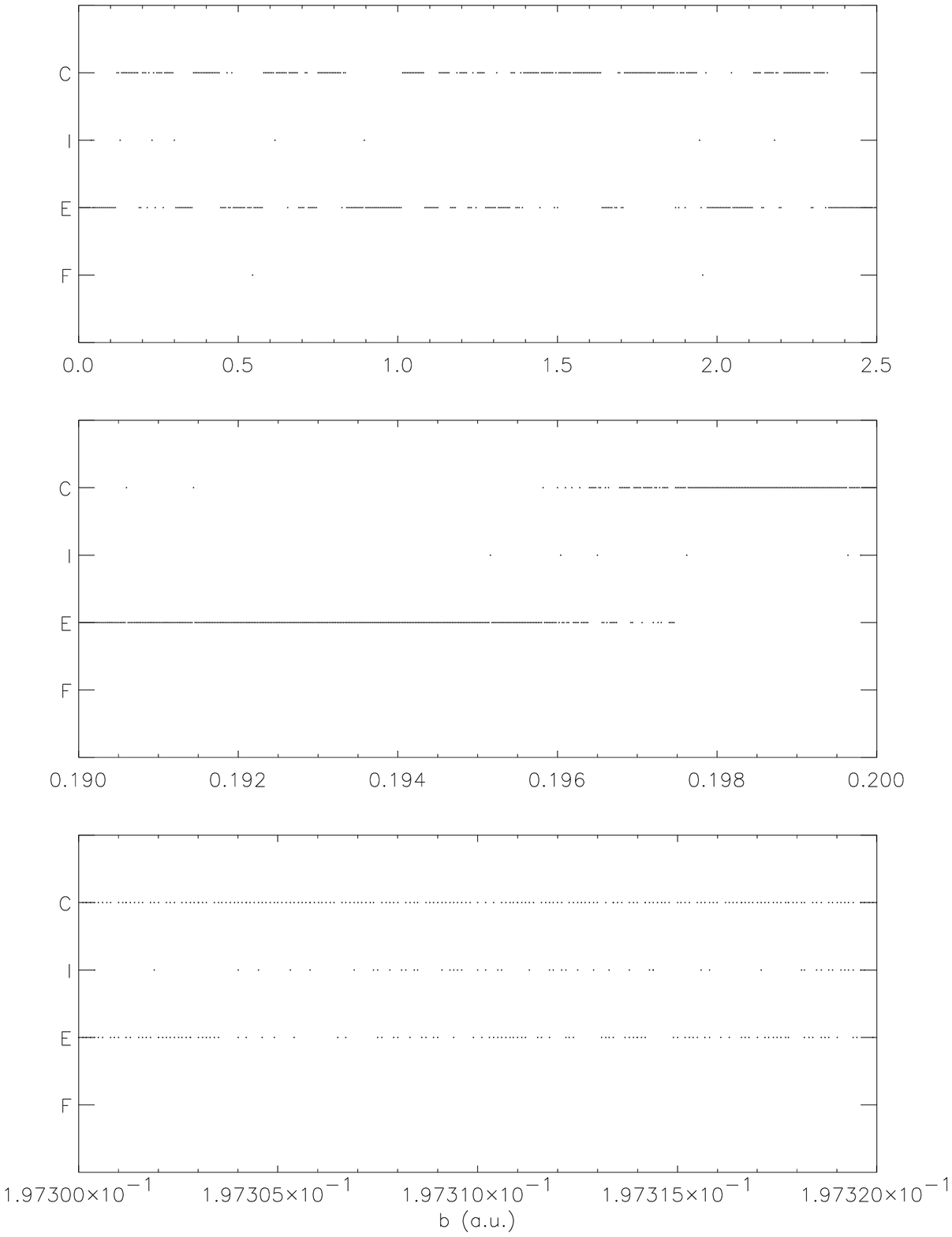}
\caption{ Scattering function {\it versus} impact parameter $b$. 'C' means 
electron capture; 'I', ionization; 'E', electronic excitation; 'F', 
undetermined.
Here $v = 0.10$, $\theta = 5.82512$ radians. From top to bottom successive 
enlargements are shown with respect to $b$. The number of runs is about 
500 for each plot.}
\label{fig1}
\end{figure}

% fig 2
\begin{figure}
\epsfxsize=12cm
\epsfbox{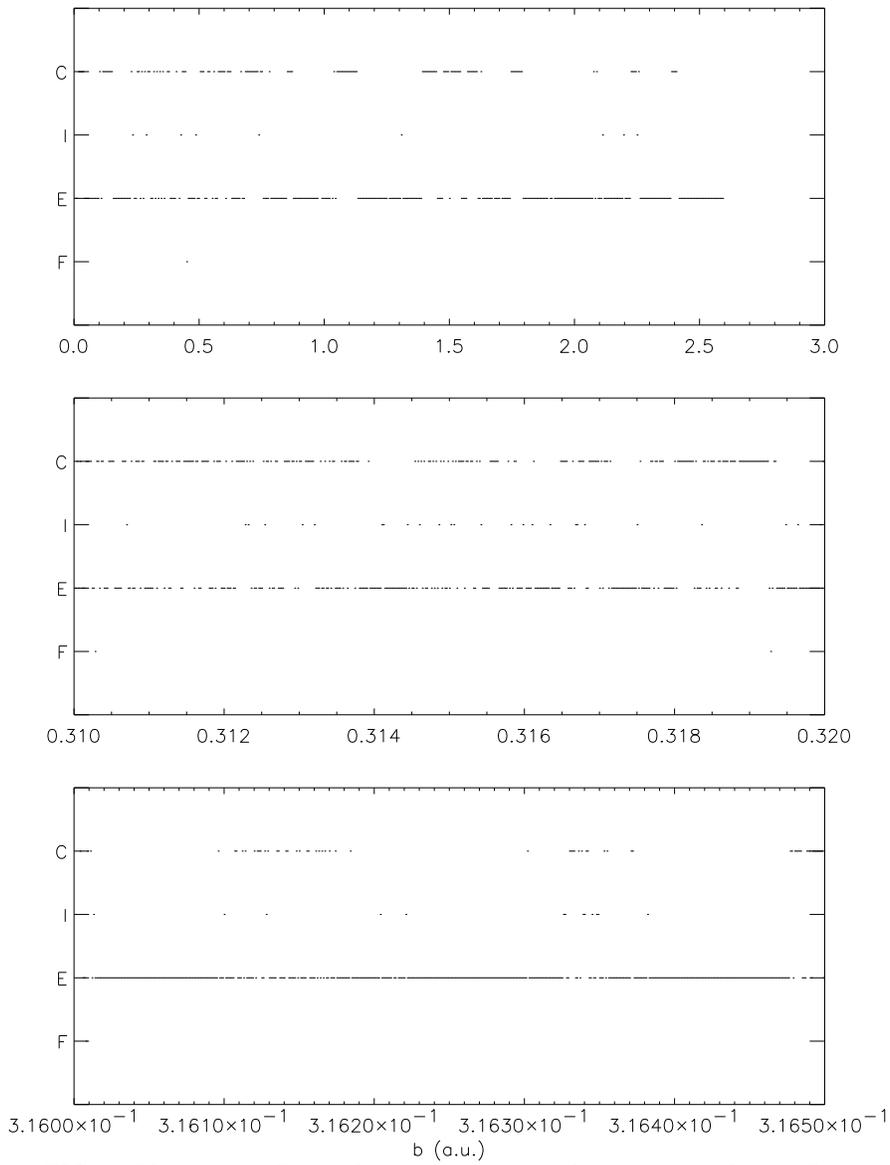}
\caption{The same as Fig. 1, but now $v = 0.120$, $\theta = 0.25262$ 
radians. The number of runs is about 500 for each plot.} 
\label{fig2}
\end{figure}

% fig 3
\begin{figure}
\epsfxsize=12cm
\epsfbox{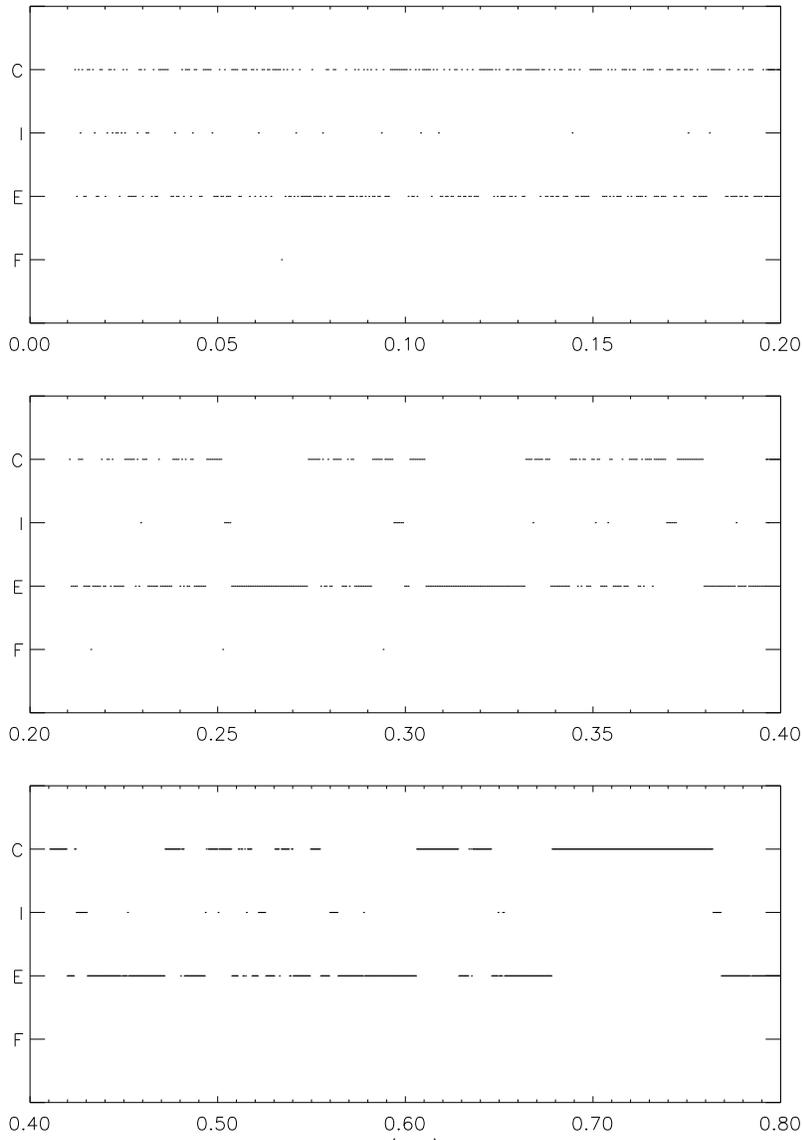}
\caption{Scattering function {\it versus} impact velocity $v$. $b = 
0.2893$, $\theta = 5.82512$ radians. The number of runs is about 400 for 
each plot.}
\label{fig3}
\end{figure}

% fig 4
\begin{figure}
\epsfxsize=12cm
\epsfbox{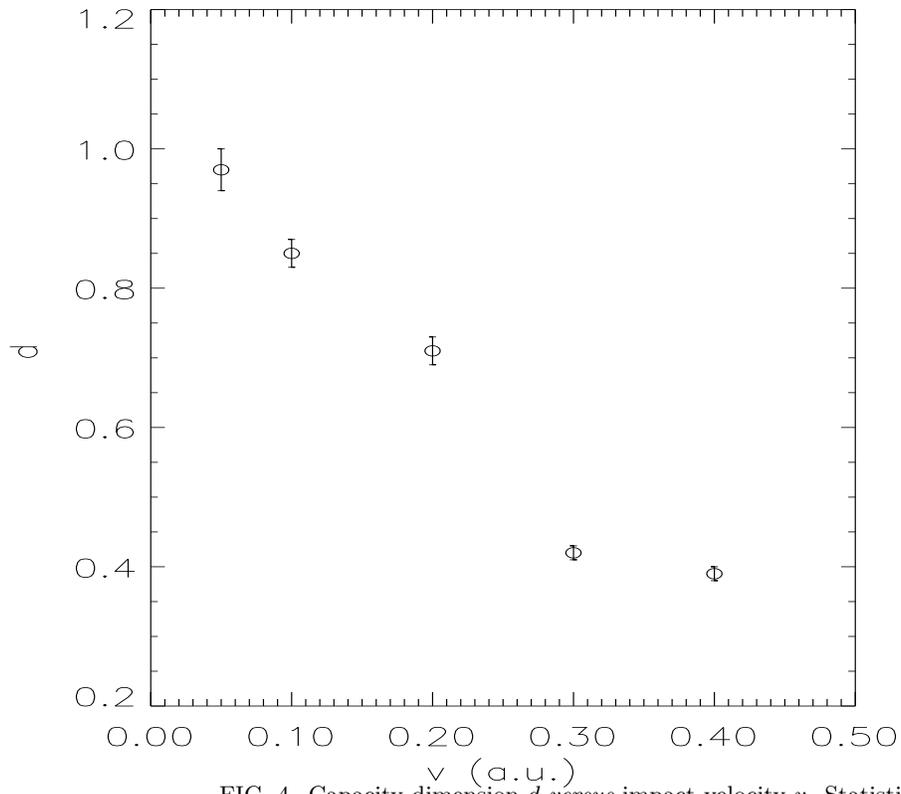}
\caption{Capacity dimension $d$ {\it versus} impact velocity $v$. 
Statistical errors are shown.}
\end{figure}

% fig 5
\begin{figure}
\epsfxsize=12cm
\epsfbox{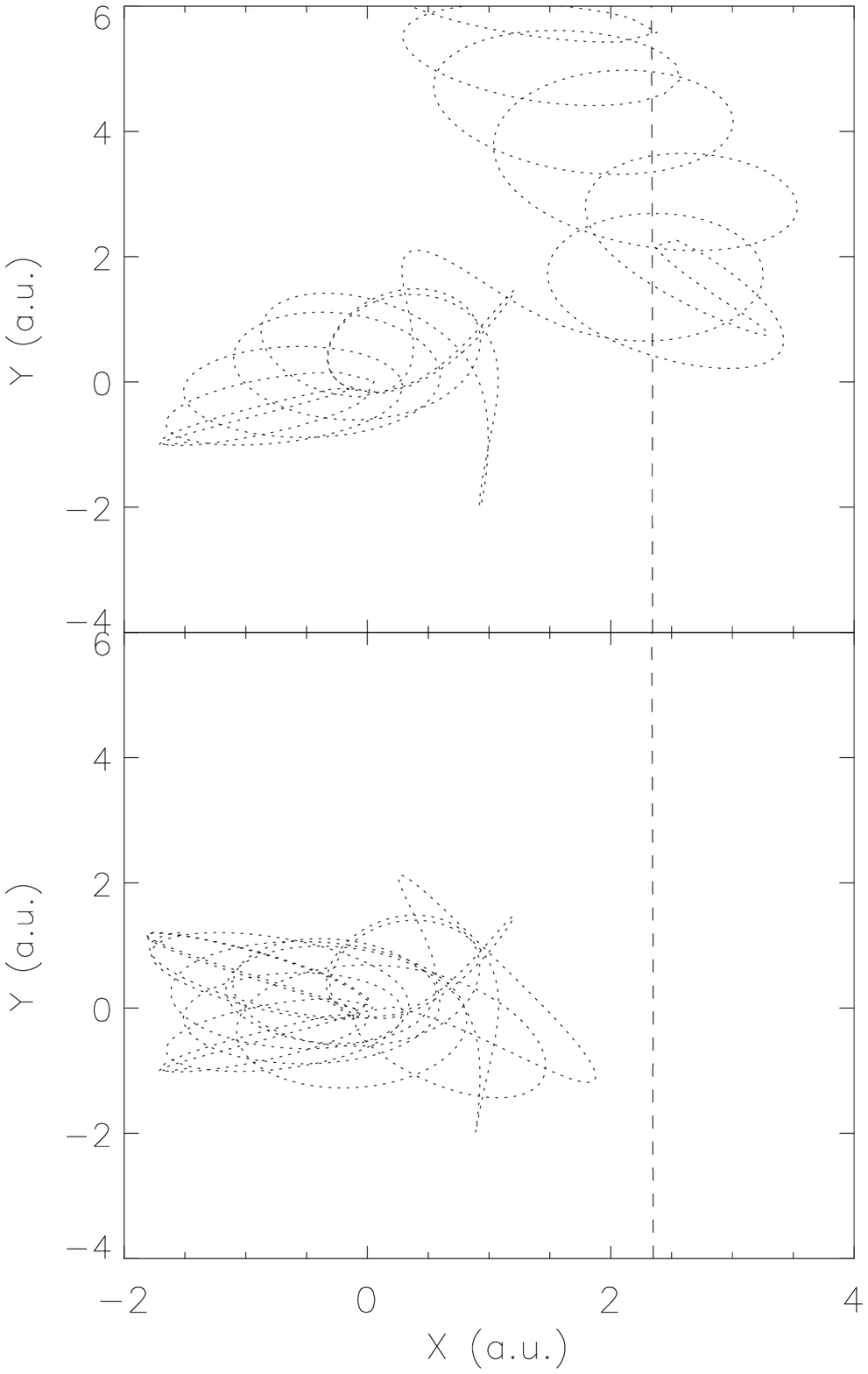}
\caption{Upper, trajectories of the particles in the interaction region. 
$v = 0.10$, $\theta = 5.82512$ radians, and $b = 2.345$. The dashed line 
is the trajectory of the heavy projectile. The electron trajectory is the 
dotted line. The initial position of the electron is close to $(0,0)$. The 
trajectory of the target nucleus is not shown since it is always close to 
the origin $(0,0)$. The lower panel shows the same process, but now with 
$b = 2.350$. } 
\label{fig5}
\end{figure}


\begin{references}

\bibitem{gaspard} P. Gaspard and S. A. Rice, J. Chem. Phys. {\bf 90}, 
2225, 2242, 2255 (1989)

\bibitem{bleher} S. Bleher, C. Grebogi, and E. Ott, Phys. Rev. Lett. {\bf 
63}, 919 (1990).

\bibitem{bleher2} S. Bleher, C. Grebogi, and E. Ott, Physica D {\bf 46}, 
87 (1990).

\bibitem{troll} G. Troll, Physica D {\bf 83}, 355 (1995). 

\bibitem{brandis} S. Brandis, Phys. Rev. E {\bf 51}, 3023 (1995). 

\bibitem{petit} J.M. Petit and M. H\'enon, Icarus {\bf 66}, 536 (1986).

\bibitem{yuan} J.-M. Yuan and Y. Gu, CHAOS {\bf 3}, 569 (1993). 

\bibitem{chaos} Focus Issue on Chaotic Scattering, T. T\'el and E. Ott 
editors, CHAOS {\bf 3} (1993).

\bibitem{kovacs} Z. Kov\'acs and L. Wiesenfeld, Phys. Rev. E {\bf 51}, 
5476 (1995).

\bibitem{sattin} F. Sattin and L. Salasnich, J. Phys. B: At. Mol. Opt. 
Phys. {\bf 29}, L699 (1996).

\bibitem{bransden} B.H. Bransden and M.R.C. McDowell, Charge Exchange and 
the Theory of Ion--Atom Collisions (Oxford Science Publications, 1992).

\bibitem{chen} Q. Chen, M. Ding and E. Ott, Phys. Lett. A {\bf 145}, 93 
(1990).

\bibitem{tokesi} K. T\"okesi and G. Hock, Nucl. Instrum. Meth. Phys. Res. 
B {\bf 86}, 201 (1994).

\bibitem{bleher1} S. Bleher, C. Grebogi, E. Ott, and R. Brown, Phys. Rev. 
A {\bf 38}, 930 (1988).

\bibitem{yamamoto} T. Yamamoto and K. Kaneko, Phys. Rev. Lett. {\bf 70}, 
1928 (1993).

\bibitem{rost1} J.-M. Rost, Phys. Rev. Lett. {\bf 72}, 1998 (1994); J. 
Phys. B: At. Mol. Opt. Phys. {\bf 27}, 5923 (1994); J. Phys. B: At. Mol. 
Opt. Phys. {\bf 28}, 3003 (1995). 

\end{references}
\end{document}